\documentclass[reprint,amsmath,amssymb,aps,prb,showpacs]{revtex4-1}
\usepackage[T1]{fontenc}

\usepackage{graphicx}
\graphicspath{ {figures/} }
\usepackage{amsmath}

\usepackage[colorlinks,hyperindex]{hyperref}
\usepackage{color}

\begin{document}

\title{Magnetic polarons in a nonequilibrium polariton condensate}

\author{Pawe{\l} Mi\k{e}tki}
\author{Micha{\l} Matuszewski}
\affiliation{Institute of Physics, Polish Academy of Sciences, Al. Lotnikow 32/46, PL-02668 Warsaw, Poland}
  
\begin{abstract}
  We consider a condensate of exciton-polaritons in a diluted magnetic semiconductor microcavity.
  Such system may exhibit magnetic self-trapping in the case of sufficiently strong
  coupling between polaritons and magnetic ions embedded in the semiconductor. We investigate the effect of the nonequilibrium nature of exciton-polaritons
  on the physics of the resulting self-trapped magnetic polarons.
  We find that multiple polarons can exist at the same time, and derive
  a critical condition for self-trapping which is different to the one predicted previously in the equilibrium case.
  Using the Bogoliubov-de Gennes approximation, we calculate the excitation spectrum and provide a physical explanation in terms of
  the effective magnetic attraction between polaritons, mediated by the ion subsystem.
\end{abstract}

\maketitle

\section{\label{sec:intro}Introduction}

Exciton-polaritons are versatile quantum quasiparticles that exist in semiconductor systems, in which the exciton-photon coupling overcomes the effects of decoherence~\cite{Polaritons}. This so-called strong coupling regime is characterized by the appearance of new branches of excitations with mixed light-matter characteristics. In semiconductor microcavities, polaritonic modes possess an effective mass orders of magnitude smaller than the electron mass, which allows for the observation of bosonic condensation even at room temperature~\cite{Kasprzak_BEC,Grandjean_RoomTempLasing,Kena_NonlinearOrganic}. This led to the observation of phenomena such as superfluidity~\cite{Amo_Superfluidity,Sanvitto_Superfluidity}, Josephson oscillations~\cite{Pietka_Josephson,Amo_Josephson}, quantum vortices~\cite{Deveaud_QuantumVortices,Sanvitto_PersistentCurrents,Deveaud_VortexDynamics}, and solitons~\cite{Amo_DarkSolitons,Sich_solitons,Sanvitto_nasz}. The applications of polaritonic condensates that are considered currently include low threshold lasers~\cite{Yamamoto_NPRev}, all-optical logic~\cite{Sanvitto_Transistor,Savvidis_TransistorSwitch,Shelykh_Neurons}, quantum simulators~\cite{Lagoudakis_XYModel,Yamamoto_QuantumSimulators} and few photon sources~\cite{Laussy_NPhotonBundles}.

Recently, exciton-polariton systems appeared as a promising platform for topological quantum states in photonic lattices~\cite{Soljacic_Review,Liew_Garden,Amo_EdgeStates,Amo_QuasicrystalEdgeStates,Malpuech_TopoReview,Malpuech_TopologicalInsulator}. Unidirectional transport in topological states can be realized by breaking time-reversal symmetry~\cite{Soljacic_Review,Malpuech_TopoReview}. In the polariton context, this is possible thanks to the exciton sensitivity to the magnetic field. However, due to the weak magnetic response, and linewidth limited by the short lifetime of polaritons, it is difficult to achieve sufficiently well resolved energy splitting of spin polarized branches~\cite{Pietka_MagneticFieldTuning}, which is a prerequisite for exploiting their topological properties. In this context, diluted magnetic (or semimagnetic) semiconductor materials appear as a promising medium for the realization of topological polariton transport. In these materials, the response to magnetic field is enhanced by orders of magnitude due to the coupling of exciton spin to the spin of magnetic ions diluted in the semiconductor medium~\cite{Kossut_Review,Dietl_Review,Ivchenko,Kavokin_CoherentSpinDMS,Mirek_AngularDependence}. Following the recent progress of sample fabrication, which led to the observation of polariton lasing in a high quality semimagnetic microcavity~\cite{Pacuski_StrongCoupling}, these systems are among the most promising in the context of realizing nontrivial topological states.

One of the most fundamental phenomena predicted in condensates of semimagnetic polaritons is the magnetic self-trapping~\cite{Kavokin_InterplaySuperfluidityDMS}, or formation of magnetic polarons~\cite{Dietl_Polaron}. It was predicted that when the ion-exciton coupling is strong enough, and at low enough temperature, self-trapping can occur, which leads to the condensation in real space and the breakdown of superfluidity~\cite{Kavokin_InterplaySuperfluidityDMS}. However, this theoretical prediction was entirely based on the equilibrium model, in which condensation in the ground state of a system without dissipation was assumed. While equilibrium condition in polariton condensates has been realized very recently in state-of-the art GaAs microcavity samples~\cite{Snoke_EquilibriumBEC}, it is not satisfied in majority of microcavities, and in particular Cd$_{1-x}$Mn$_x$Te systems, which possess strong magnetic properties. It is therefore important to investigate the effect of nonequilibrium nature of polariton condensates on the existence and properties of magnetic polarons.

In this paper, we investigate in detail magnetic self-trapping in polariton condensates, while fully taking into account the nonequilibrium physics of the system. At the same time, we assume that the magnetic ion subsystem is fully thermalized as evidenced in experiments~\cite{Mirek_AngularDependence}. We find that in the nonequilibrium case, multiple magnetic polarons can be formed at the same time, in contrast to previous findings~\cite{Kavokin_InterplaySuperfluidityDMS}. We investigate both the case of homogeneous pumping with periodic boundary conditions, and a more realistic case of Gaussian pumping. Moreover, we find that the critical condition for self-trapping differs from the one predicted in equilibrium, as the polariton temperature cannot be defined. We obtain diagrams of stability in function of the ion-polariton coupling, temperature, and magnetic field. Additionally, we use the Bogoliubov-de Gennes approximation to examine the stability of a uniform condensate against self-trapping. We derive an analytic formula for the stability threshold, which, surprisingly, does not depend on the spin relaxation time of the magnetic ions. These results are confirmed numerically and explained by the effective nonlinearity in the model. The self-trapping is directly connected to the effective magnetic attraction between polaritons, induced by the coupling to the ion subsystem. 

\section{\label{sec:model}Model}

We consider an exciton-polariton condensate in a two-dimensional semimagnetic semiconductor microcavity. The cavity contains quantum wells that are composed of a diluted magnetic semiconductor (such as Cd$_{1-x}$Mn$_x$Te) with incorporated magnetic ions, a setup that has been realized recently~\cite{Pacuski_StrongCoupling,Mirek_AngularDependence}.
We consider a specific case of a two-dimensional cavity where polaritons are confined in a one-dimensional geometry by a microwire or line defect~\cite{Bloch_ExtendedCondensates,Deveaud_Disorder1D}. In this work we will assume that the condensate is fully spin-polarized. This can be achieved experimentally by the combined effect of a circularly polarized pump, and the influence of external magnetic field in the direction perpendicular to the quantum well, which suppresses exciton spin-flip. We also neglect the effects of TE-TM splitting which could lead to the precession of polariton spins in an effective magnetic field~\cite{Shelykh_PolarizationPropagation}.

In the case of tight transverse confinement, the evolution of the condensate can be described by the complex Ginzburg-Landau equation (CGLE) coupled to the equation describing spin relaxation of magnetic ions~\cite{Kavokin_Polaron}
\begin{gather}
\begin{aligned}
i \hbar \frac{\partial \psi}{\partial t} = - \frac{\hbar^2}{2 m^*} &\frac{\partial^2 \psi}{\partial x^2} + g_{\rm C} |\psi|^2 \psi - \gamma_{\rm NL} |\psi|^2 \psi - \lambda M \psi \\
&+ i P(x) \psi - i\frac{1}{2}\gamma_{\rm L} \psi \label{eq:CGLE1}
\end{aligned} \\
\frac{\partial M(x,t)}{\partial t} = \frac{ \langle M(x, t) \rangle - M(x,t)}{\tau_{\rm M}} \label{eq:MTR1}
\end{gather} 
where $g_{\rm C}$ is the polariton interaction constant, $\gamma_{\rm L}$ and $\gamma_{\rm NL}$ are linear and non-linear loss coefficients, respectively, $\lambda$ is the magnetic ion-polariton coupling constant, $\tau_{\rm M}$ is the spin relaxation time of magnetic ions, and $P(x)$ is the space-dependent external pumping rate. We note that the nonlinear coefficients have been rescaled in the 1D case and are related to their 2D counterparts through  $(g_{\rm C}^{1D}, \gamma_{\rm NL}^{1D}) = (g_{\rm C}^{2D}, \gamma_{\rm NL}^{2D}) /\sqrt{2\pi d^2}$, where $d$ is the lengthscale of the transverse confinement.
Here, we assumed a Gaussian transverse profile of $|\psi|^2$ of width $d$.
In the case of a one-dimensional microwire~\cite{Bloch_ExtendedCondensates}, 
the profile width $d$ is of the order of the microwire thickness.
We emphasize that the pumping and loss terms in Eq.~(\ref{eq:CGLE1}) were absent in the previous study of magnetic self-trapping~\cite{Kavokin_InterplaySuperfluidityDMS}.

The equilibrium ion magnetization in the dilute regime is given by the Brillouin function~\cite{Gaj_Brilloiun}
\begin{gather}
	\langle M(x, t) \rangle= n_{\rm M}g_{\rm M} \mu_{\rm B} J B_J\left( \frac{g_{\rm M} \mu_{\rm B} J B_{\rm eff}}{k_{\rm B} T} \right)
\end{gather}
where $n_{\rm M}$ and $g_{\rm M}$ are the 1D concentration and \textit{g}-factor of magnetic ions with total spin $J=5/2$,
$\mu_{\rm B}$ is the Bohr magneton, $T$ is the ion subsystem temperature, and $B_{\rm eff}$ is an effective magnetic field that consists of an external magnetic field $B_0$ and a contribution from the interaction with the polarized condensate
\begin{gather}
	B_{\rm eff} = B_0 + \frac{1}{2} \lambda |\psi|^2 = B_0 + \lambda S_z, \label{Beff1}
\end{gather}
where $S_z$ is the polariton $1/2$-pseudospin density. Here, because of the assumption of full condensate polarization, $S_z$ is simply equal to half of the polariton density $n=|\psi|^2/2$.
The coupling constant $\lambda$ can be estimated as~\cite{Kavokin_InterplaySuperfluidityDMS}
\begin{gather}
	\lambda = \frac{\beta_{ex} X^2}{\mu_{\rm B} g_{\rm M} L_z} 
\end{gather}
where $\beta_{ex}$ is the ion-exciton exchange interaction constant, $X$ is the excitonic Hopfield coefficient, and $L_{\rm z}$ is the width of the quantum well.

We consider two cases of space dependence of the pumping profile $P(x)$, i.e. homogeneous and Gaussian pumping.
In the case of uniform pumping the effective pumping is simply the difference of pumping and linear loss terms, $P_{\rm eff} = P - \gamma_{\rm L}$.
In the Gaussian pumping case we assume
\begin{gather}
     P(x) = P_1 \exp \left( - \frac{x^2}{2 \sigma_p^2} \right).
\end{gather}
where $\sigma_p$ corresponds to the spatial width of the pump beam.

\section{\label{sec:num_results}Numerical results}

\begin{figure}[b]
	\includegraphics[width=0.5\textwidth]{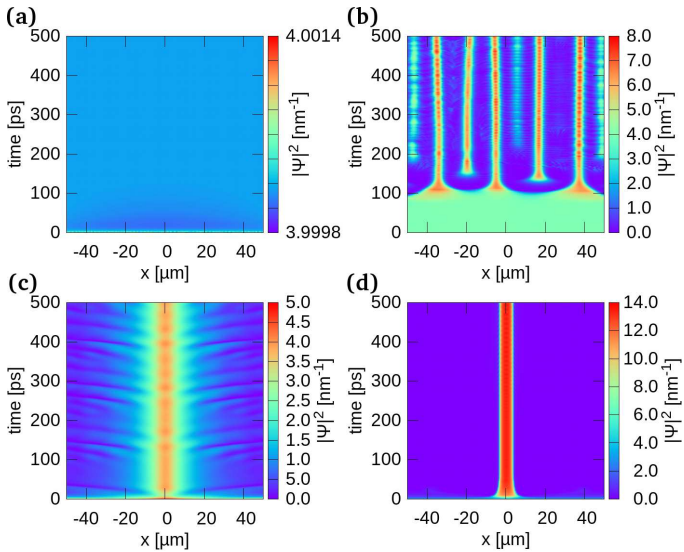}    
    \caption{
    Evolution of the norm $|\psi|^2$ of the condensate wavefunction. 
    Upper plots show the case of uniform pumping; 
    bottom plots show the case of Gaussian pumping.
    Left-hand plots show stable cases when $\lambda < \lambda_C$;
    right-hand plots show unstable cases in which polarons are formed.
    Parameters:
    $\gamma_{\rm L}$=$\mathrm{ 0~meV}$ in (a),(b);
    $\gamma_{\rm L}$=$\mathrm{ 6.582 \times 10^{-2}~meV}$ in (c),(d);
    $\lambda$=$\mathrm{ 7.125 \times 10^{-12}~T~m }$ in (a),(c)
    $\lambda$=$\mathrm{ 8.125 \times 10^{-12}~T~m}$ in (b); 
    $\lambda$=$\mathrm{ 9.5 \times 10^{-12}~T~m}$ in (d).
    Others parameters are given in \cite{fig1 parameters}}.
    \label{fig:evolution}
\end{figure}

\begin{figure}[t]
	\includegraphics[width=0.5\textwidth]{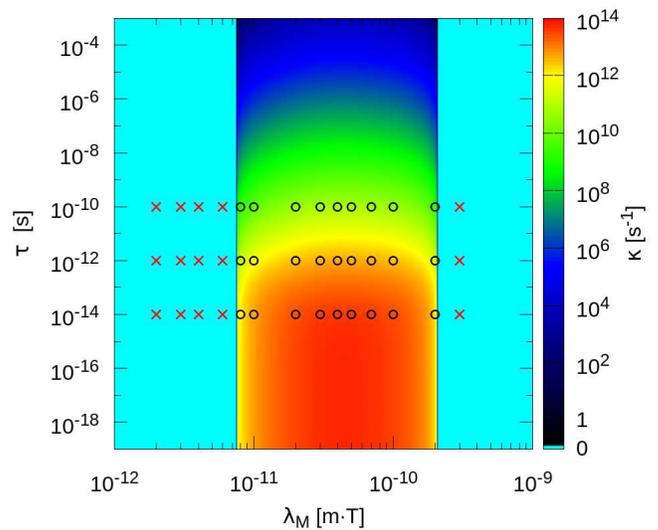}
    \caption{
    Diagram of stability shown in coordinates of the ion-polariton coupling constant $\lambda$ and the spin relaxation time of magnetic ions $\tau$.
	Stability limits were calculated analytically (see Sec.~\ref{sec:BdGapprox}).
    Color scale represents the instability rate according to the Bogoliubov-de Gennes approximation;
    Cyan color shows that the system is stable (it is symbolically expressed by ``$0$'' on the logarithmic scale);
    Circles correspond to unstable states as predicted by the simulation of Eqs.~(\ref{eq:CGLE1})-(\ref{eq:MTR1}); 
    Crosses correspond to stable states.}
    \label{fig:tau-lambda}
\end{figure}
\begin{figure}[t]
 	\includegraphics[width=0.5\textwidth]{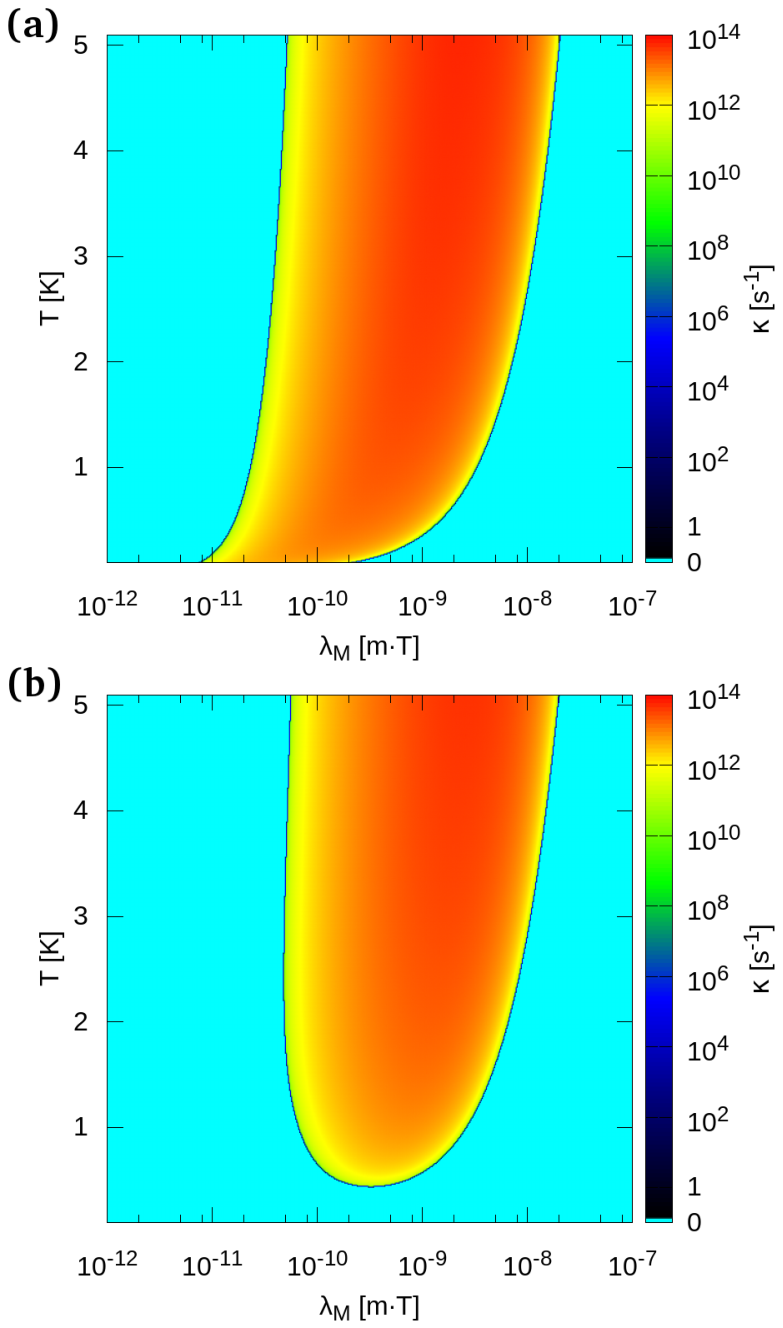}
    \caption{
    Diagrams of stability shown in coordinates of the ion-polariton coupling constant $\lambda$ and temperature $T$.
    (a) The case without magnetic field;
    (b) the case with magnetic field of $B=1$ T. 
    Color code and parameters are the same as in Fig. \ref{fig:tau-lambda} and the spin relaxation time is $\tau=10^{-12}$~s.}
    \label{fig:T-lambda-2}
\end{figure}

It was demonstrated~\cite{Kavokin_InterplaySuperfluidityDMS} that when the coupling between the polariton and magnetic subsystems is strong enough, self-trapping can occur due to the magnetic polaron effect~\cite{Dietl_Polaron,Kavokin_Polaron}. The critical condition for self-trapping was given under the assumption of thermal equilibrium in the system. We note, however, that in the majority of current experiments thermal equilibrium is not achieved. In contrast, condensation of exciton-polaritons often takes place in far from equilibrium conditions, where it is driven by the system kinetics without a well defined temperature of the polariton subsystem. It is therefore important to investigate what is the influence of the nonequilibrium character of polariton condensation on the magnetic polaron effect.

Before analyzing the precise conditions for self-trapping, we demonstrate examples typical behavior of the system. In Fig.~\ref{fig:evolution}, we show examples of dynamics obtained from Eqs.(\ref{eq:CGLE1}) and (\ref{eq:MTR1}) under both uniform and Gaussian pumping and at $B=0$. 
The numerical space window was set to be $x \in (-100\mu$m$,100 \mu$m$)$ with periodic boundary conditions, and the parameters correspond to a Cd$_{1-x}$Mn$_x$Te sample with a few percent concentration of Mn ions. In the case of Gaussian pumping,
absorbing boundary conditions were implemented at the edges of the numerical grid.
In all cases the initial state was given by a homogeneous state $(n = |\psi_0|^2)$ close to the stationary state of model equations, perturbed by a small white Gaussian noise.

Figures \ref{fig:evolution}(a) and \ref{fig:evolution}(b) correspond to the case of homogeneous pumping. At a lower value of the ion-exciton coupling constant $\lambda$, the homogeneous state given by the stationary condition $n = P_0/\gamma_{\rm NL}$ is stable, as shown in Fig.~\ref{fig:evolution}(a). However, when the coupling constant becomes higher than a certain threshold $\lambda_{\rm c}$, formation of localized polarons can be observed in Fig.~\ref{fig:evolution}(b). The polarons are almost stationary and surrounded by areas of very low polariton density. Some temporal oscillations of polaron widths can be seen. Polarons are characterized by both high polariton density and ion magnetization (not shown), which evidences the interaction between these subsystems.

In the case of a Gaussian pumping profile, below critical threshold for self-trapping a condensate is formed in the area covered by the pumping beam. As shown Fig. \ref{fig:evolution}(c), it may also exhibit oscillations which are however not related to the polaron effect. Crossing the threshold $\lambda_{\rm c}$ leads to a dramatic reduction of the spatial size of the condensate, as shown in Fig. \ref{fig:evolution}(d), in agreement with results obtained in~\cite{Kavokin_InterplaySuperfluidityDMS}.

The above examples are generic and correspond to dynamics occurring generally at various values of model parameters. Therefore we conclude that the polaron self-trapping effect can be observed in a nonequilibrium condensate, although in contrast to previous study~\cite{Kavokin_InterplaySuperfluidityDMS}, we find that multiple polarons can exist in the system at the same time, both in the case of homogeneous and Gaussian pumping. We investigated the parameter space of the model in a systematic way to determine the conditions for self-trapping in a nonequilibrium system.
The phase diagram in the space of coupling constant and relaxation time for homogeneous pumping is shown in Fig.~\ref{fig:tau-lambda}. Results of numerical simulations of model equations are indicated by crosses (stable condensate) and circles (self-trapping). Additionally, we show the results of Bogoliubov-de Gennes analysis of stability of the uniform state (see Sec.~\ref{sec:BdGapprox} for details), which are given by the color scale. Clearly there is a very good agreement between the analytical predictions and the results of numerical simulations. 
One can observe that the coupling constant $\lambda$ is the most important parameter that determines the stability and $\tau_{\rm M}$ has only the influence on the instability rate of the steady state, which corresponds to the time necessary for the formation of polarons. 

Figure~\ref{fig:T-lambda-2} contains phase diagrams (according to the BdG stability analysis) in the space of the ion temperature $T$ and ion-polariton coupling $\lambda$, at (a) zero magnetic field and (b) magnetic field of 1T. Note that (a) suggests that in the case of low coupling constant, very low temperatures are necessary to observe the polaron effect. However, this dependence becomes less pronounced for higher values of $\lambda$.
The main effect of the magnetic field exists at small temperatures, where the uniform condensate becomes stable for all values of $\lambda$.

Finally, we note that the degree of circular polarization of the condensate depends, among other parameters, on the temperature and magnetic field, due to the existence of spin-flip  processes. For this reason, in particular in the $B=0$ case, condensate can become polarized elliptically or linearly at low temperatures, which will lead to the modification of the phase diagram shown in Fig. \ref{fig:T-lambda-2}a.
\section{\label{sec:BdGapprox}Stability analysis}

In this section we apply the Bogoliubov-de Gennes approximation in the case of uniform pumping to find an analytical condition of stability of a stationary state. We postulate that the emergence of magnetic polarons corresponds to the instability threshold for the uniform state. Indeed, we find a full analogy of the effective nonlinearity emerging from the model Eqs.~(\ref{eq:CGLE1})-(\ref{eq:MTR1}) in the fast relaxation rate regime to the Gross-Pitaevskii equation with attractive nonlinearity. In terms of this correspondence, polarons can be identified as bright solitons emerging from an unstable uniform background~\cite{Kivshar_OpticalSolitons}.

For the sake of clarity of the derivation we now introduce a dimensionless form of the model.
Equations (\ref{eq:CGLE1}),(\ref{eq:MTR1}) can be transformed by rescaling time, space, wavefunction amplitude, and system parameters as $t = \alpha \tilde{t}$, $x =\xi \tilde{x}$, $\psi = (\xi\beta)^{-1/2} \tilde{\psi}$, $g_{\rm C} = \hbar\xi\beta\alpha^{-1} \tilde{g_{\rm C}}$, $P_{\rm eff} = \hbar\alpha^{-1} \tilde{P}$, $\gamma_{\rm NL} = \hbar\xi\beta\alpha^{-1} \tilde{\gamma_{\rm NL}}$, $M = \zeta \tilde{M}$, $\lambda = \hbar\alpha^{-1} \tilde{\lambda}$ to obtain the dimensionless form (hereafter we omit the tildes)
\begin{gather}
i \frac{\partial \psi}{\partial t} = - \frac{\partial^2 \psi}{\partial x^2} + g_{\rm C} |\psi|^2 \psi + iP \psi - i\gamma_{\rm NL} |\psi|^2 \psi - \zeta \lambda M \psi \label{CGLE2} \\
\frac{\partial M}{\partial t} = \frac{\alpha}{\tau_{\rm M}} \left[J B_J\left( \delta \lambda |\psi|^2 \right)  - M \right] \label{MTr2} 
\end{gather}
where $\xi = \sqrt{\hbar\alpha/2m^*}$, $\zeta = g_{\rm M} \mu_{\rm B} n_{\rm M}$, $\delta = \frac{g_{\rm M} \mu_{\rm B} J}{2 k_{\rm B} T} \frac{\hbar}{\alpha \beta \xi }$ and $\alpha$, $\beta$ are free parameters. 

Fluctuation around the stationary homogeneous solution $\psi_0(x,t) = n^{1/2} e^{-i \mu t}$, $M_0(x,t) = J B_J \left( \delta \lambda | \psi_0 |^2 \right)$ can be written in the plane wave basis~\cite{Wouters_ExcitationSpectrum} (note that $n = P/\gamma_{\rm NL}$)
\begin{gather}
	\psi(x,t) = n^{1/2} e^{-i \mu t} \left[1 + \epsilon \sum_k \left\{ u_k(t) e^{ikx} + v_k(t) e^{-ikx} \right\} \right] \\
    M(x,t) = M_0 + \epsilon \sum_k \left[ w_k(t) e^{ikx} + w_k^*(t) e^{-ikx} \right]
\end{gather}
where $\epsilon$ is a small perturbation parameter.

The linearized solution is obtained by taking $\epsilon$ up to the first order, expanding Brillouin function about $\psi_0$ up to the first term and comparing parts with $e^{ikx}$ and $e^{-ikx}$ respectively.
It can be rewritten as the following eigenvalue problem
\begin{align} \label{eq:eigenvalue}
 i \frac{\mathrm{d}}{\mathrm{d}t} 
 \begin{pmatrix} u \\ v^* \\ w \end{pmatrix} = Q
 \begin{pmatrix} u \\ v^* \\ w \end{pmatrix}
\end{align}
where the matrix $Q$ is given by
\begin{align}
	Q = 
\begin{pmatrix} 
k^2 -iEn + g_{\rm C}n & -iEn + g_{\rm C}n       & - \zeta \lambda n^{1/2} \\
-iEn - g_{\rm C}n     & -k^2 -iEn - g_{\rm C}n  &  \zeta \lambda n^{1/2} \\
i\frac{\alpha}{\tau} \delta \lambda n^{1/2} J B_J^\prime( \delta \lambda n) &
i\frac{\alpha}{\tau} \delta \lambda n^{1/2} J B_J^\prime( \delta \lambda n) & 
-i\frac{\alpha}{\tau}
\end{pmatrix}
\label{eq:Qmatrix}
\end{align}

The numerical solution of the above eigenvalue problem in parameter space is shown in Fig. \ref{fig:tau-lambda} and Fig. \ref{fig:T-lambda-2} in color scale, which corresponds to the most unstable mode (highest imaginary part of the eigenfrequency) of the system~(\ref{eq:eigenvalue}). Parameters with stable evolution (all eigenvalues with zero or negative imaginary part) are depicted with cyan color.

Additionally, it is possible to derive an exact analytical condition for the stability of the system. The procedure is analogous to the one described in~\cite{Ostrovskaya_DarkSolitons} and consists of the analysis of the zero-frequency crossing of the imaginary part of the eigenfrequency in function of momentum $k$. The existence of the crossing indicates that momenta with eigenfrequencies with both positive and negative imaginary parts exist on two sides of the crossing. 
The eigenvalue problem of Eq. (\ref{eq:Qmatrix}) leads to the equation for $\omega(k)$
\begin{gather}
	\omega^3 + i(Y+2R)\omega^2 - (\omega_B^2+2YR)\omega = iY(\omega_B^2 - 2Gk^2)
    \label{eq:dispersion}
\end{gather}
where $\omega_B$, $Y$, $R$ and $G$ are defined by
\begin{gather}
    \omega_B^2 = k^4 + 2k^2 g_{\rm C}^{ }n  \\
    Y = \frac{\alpha}{\tau},~~R = En, ~~G = \zeta\delta\lambda^2J B_J^\prime( \delta \lambda n)n
\end{gather}
We can analyze the solutions in the limits $k \rightarrow \infty$ and $k \rightarrow 0$.
In the $k \rightarrow \infty $ limit, there are three branches: $\omega \approx -iY , \pm k^2 - iR$, and all have negative imaginary parts.
In the $k \rightarrow 0$ limit, there are two  solutions with negative imaginary parts and one equal to zero: $\omega_1(0) = 0$, $\omega_2(0) = -iY$ and $\omega_3(0) = - 2iR$.
Only the $\omega_1$ branch can cross the zero-frequency axis and have positive imaginary part in some range of $k$.
The crossing points can be found by putting $\Im (\omega) = 0$ and $\Re (\omega) = \Omega$ into Eq. (\ref{eq:dispersion}), which have to be satisfied at the same time
\begin{gather}
	\label{eq:omega1}
	\Omega\left(\Omega^2 - (\omega_B^2+2YR)\right) = 0 \\
	(Y+2R)\Omega - Y\omega_B^2 + 2YGk^2 = 0
    \label{eq:omega2}
\end{gather}
For physical parameters Eqs.(\ref{eq:omega1}),(\ref{eq:omega2}) are realized only if $\Omega=0$. That leads to the equation for $k$ 
\begin{gather}
	k^4 + 2k^2 g_{\rm C}^{ }n  + 2 Gk^2 = 0
\end{gather}
and the analytical condition for stability, which expressed in physical units reads
\begin{gather}
	\lambda^2 B_J^\prime \left( \frac{g_{\rm M} \mu_{\rm B}}{2 k_{\rm B} T} \lambda n J\right) < \frac{2 g_{\rm C} k_{\rm B} T}{n_{\rm M} g_{\rm M}^2 \mu_{\rm B}^2 J^2} \label{eq:condition}
\end{gather} 
With respect to $\lambda$, the inequality (\ref{eq:condition}) is not satisfied in the interval $\lambda_{\rm c1}<\lambda< \lambda_{\rm c2}$, as can be seen in Fig,~\ref{fig:tau-lambda}. 
Remarkably, the critical values of $\lambda$ do not depend on the spin relaxation time $\tau$. This conclusion is fully supported by the numerical simulations as shown in Fig. \ref{fig:tau-lambda}. In this Figure, the color scale shows the calculated largest eigenvalue of the unstable branch, $\kappa={\rm max}({\rm Im}\,\omega_1(k))$. Parameters for which ${\rm Im}\,\omega_1(k)$ is always non-positive are marked with cyan color.

\subsection{Adiabatic regime}

As is shown in Fig.~\ref{fig:tau-lambda}, if the ion spin relaxation time $\tau$ is shorter than $10^{-14}\,$s, the instability rate no longer depends on the value of $\tau$. In this adiabatic regime, the fast spin relaxation approximation can be applied, which corresponds to setting the time derivative on the left hand side of Eq.~(\ref{MTr2}) to zero. In this limit we have $M(x,t) = \langle M(x, t) \rangle = J B_J\left( \delta \lambda |\psi|^2 \right)$, which gives
\begin{gather}
\begin{aligned}
i \frac{\partial \psi}{\partial t} = - \frac{\partial^2 \psi}{\partial x^2} 
&+ \left( g_{\rm C} |\psi|^2  - \zeta\lambda\langle M(x, t) \rangle \right) \psi \\
&-i\left( \gamma_{\rm NL} |\psi|^2 - P \right) \psi.
\end{aligned}
\end{gather}
Simple and intuitive interpretation of the instability can be obtained if the Brillouin function is expanded up to the first order around the stationary value of $|\psi_0|^2$
\begin{gather}
\begin{aligned}
M(x,t) = J B_J (\delta \lambda |\psi|^2) \approx J B_J (\delta \lambda |\psi_0|^2 ) \\
+ J \delta \lambda ( |\psi|^2 - |\psi_0|^2 ) B_J^\prime ( \delta \lambda |\psi_0|^2 )
\end{aligned}
\end{gather}
which leads to the standard form of the complex Ginzburg-Landau equation (or dissipative Gross-Pitaevskii equation)
\begin{gather}
\begin{aligned}
i \frac{\partial \psi}{\partial t} = - \frac{\partial^2 \psi}{\partial x^2} 
&+ \left[ \left(g_{\rm C} - \zeta\lambda^2\delta B'_J\right) |\psi|^2  - \zeta\lambda U_0 \right] \psi \\
&-i\left( \gamma_{\rm NL} |\psi|^2 - P \right) \psi.
\end{aligned}
\end{gather}
where $U_0=J B_J (\delta \lambda |\psi_0|^2 ) - J \delta \lambda|\psi_0|^2 B_J^\prime ( \delta \lambda |\psi_0|^2 )$ and the notation $B'_J=B'_J(\delta \lambda |\psi_0|^2)$ was used. The above form corresponds to the complex Ginzburg-Landau equation with effective nonlinearity
\begin{equation}
  g_{\rm eff} = g_{\rm C} - J\zeta\lambda^2\delta B'_J,
\end{equation}
which becomes attractive exactly at the threshold given by Eq.~(\ref{eq:condition}) when expressed in physical units. In other words, the instability threshold that marks the formation of polarons corresponds to the Benjamin-Feir-Newell criterion of stability of the CGLE equation~\cite{Aranson_CGLEWorld}.

\begin{figure}[t]
	\includegraphics[width=0.4\textwidth]{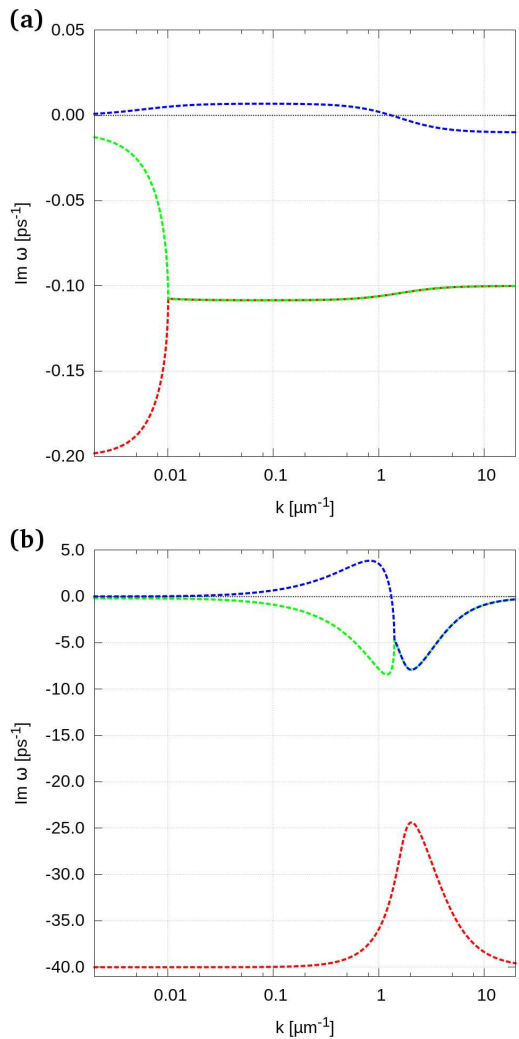}
    \caption{Imaginary parts of the frequencies of Bogoliubov quasiparticles. The ion-polariton coupling constant $\lambda$ is $10^{-11}$ T~m.
    (a) A typical excitation spectrum; spin relaxation time $\tau$ = $10^{-10}s$. The unstable branch exhibits a positive imaginary part. (b) Large instability rate at $\tau$ = $2.5\times10^{-14}$s.}
    \label{fig:dispersion}
\end{figure}
\subsection{Quasiparticle spectrum}

Figure~\ref{fig:dispersion} shows the imaginary part of the excitation spectrum of the uniformly pumped condensate. 
Figure~\ref{fig:dispersion}(a) is an example of a weak instability in the case of long ion spin relaxation rate, which corresponds to circles in the green area in Fig. \ref{fig:tau-lambda}. The spectrum contains three branches, of which one (blue line) is unstable at certain wavevector range, which is indicated by the positive imaginary part of the frequency. We note that the spectrum in this regime is strikingly similar to the one predicted in the case of a nonmagnetic condensate in the presence of a reservoir~\cite{Wouters_ExcitationSpectrum,Bobrovska_Stability}. The similarity indicates that the magnetic ions play in some sense the role of reservoir in this system.

On the other hand, in the case of a short spin relaxation time the spectrum becomes qualitatively different, as indicated in Fig.~\ref{fig:dispersion}(b). These parameters correspond to circles in the orange area of  Fig.~\ref{fig:tau-lambda}. While also one of the branches is unstable, it is strongly peaked at high momenta. Moreover, the instability rate, defined as the maximum value of the imaginary part of the frequency, is much higher than in the previous case. This indicates clearly that the relaxation time determines the timescale of the instability, as demonstrated in Fig.~\ref{fig:tau-lambda}. The lower branch (red line) has been pushed down to very low imaginary frequencies, which is characteristic of a strongly damped mode. This damped mode is the one which corresponds to the excitation of magnetization, which is strongly suppressed in the adiabatic regime.

\section{\label{sec:conclusions}Conclusions}

In conclusion, we investigated a spin-polarized condensate of exciton-polaritons in a diluted magnetic semiconductor microcavity.
In contrast to previous works, we included nonequilibrium effects of driving and decay in our model, which led to several interesting effects. We found that multiple polarons can exist at the same time, and connected the instability of the homogeneous state in the Bogoliubov-de Gennes approximation to the formation of polarons. We derived
a critical condition for self-trapping which is different to the one predicted previously in the equilibrium case. The effect has been explained by the effective attraction between polaritons due to the magnetic ion coupling.

\acknowledgements

We acknowledge support from the National Science Center grants 2015/17/B/ST3/02273 and 2016/22/E/ST3/00045.


\begin{thebibliography}{99}
\bibitem{fig1 parameters}
    $n$ = $\mathrm{4 \times 10^9}$ $\mathrm{m^{-1}}$, 
    $\tau$ = $\mathrm{ 10^{-12}}$ $\mathrm{ps }$, 
    $g_{\rm C}$ = $\mathrm{ 1.2 \times 10^{-9}}$ $\mathrm{meV~m}$, 
    $n_{\rm M}$ = $\mathrm{ 10^{13}}$ $\mathrm{m^{-1}}$, 
    $B$ = $\mathrm{ 0~T }$, 
    $T$ = $\mathrm{ 0.1~K}$, 
    $\gamma_{\rm NL}$ = $\mathrm{ 1.646 \times 10^{-11}}$ $\mathrm{meV~m}$, 
    $P_0$ = $\mathrm{ 0.0658~meV}$, 
    $P_1$ = $\mathrm{ 0.2633~meV}$, 
    $\sigma_p $ = $\mathrm{ 3.617 \times 10^{-6}}$ $\mathrm{m}$,
    $d$ = $\mathrm{8 \times 10^{-7}}$ $\mathrm{m}$.
	There was 5 meV Rabi splitting and zero photon-exciton detuning;
    photon effective mass $m_{\rm C}$ = $\mathrm{10^{-5}}~ m_{\rm E}$
    and exciton effective mass $m_{\rm X}$ = $0.22~ m_{\rm E}$.
\end{thebibliography}
\end{document}